\UseRawInputEncoding
\documentclass[10.5pt,a4paper]{article}

\usepackage[dvipdfmx]{graphicx}
\usepackage{subfigure}
\usepackage{here}

\usepackage{enumerate}
\usepackage{paralist}

\makeatletter
\def\Hline{\noalign{\ifnum0=`}\fi\hrule\@height 3.\arrayrulewidth \futurelet\reserved@a\@xhline}
\makeatother

\usepackage[top=30truemm,bottom=30truemm,left=20truemm,right=20truemm]{geometry}
\usepackage{fancyhdr}
\usepackage{amsmath}
\usepackage{amssymb}
\usepackage{mathcomp}
\usepackage{siunitx} 
\usepackage{cases}

\newcommand{\diff}{\mathrm{d}} 
\newcommand{\ai}{\mathrm{i}} 

\usepackage{authblk} 
\usepackage{color}

\usepackage{comment}

\title{Can the phase of radiation pressure fluctuations be flipped in a single path for laser interferometric gravitational wave detectors?}

\author[1,$\dagger$]{Tomohiro Ishikawa}
\author[1]{Shoki Iwaguchi}
\author[1]{Bin Wu}
\author[1]{Izumi Watanabe}
\author[1]{Yuki Kawasaki}
\author[1]{Ryuma Shimizu}
\author[2]{Yutaro Enomoto}
\author[3]{Yuta Michimura}
\author[2,4]{Akira Furusawa}
\author[1,5]{Seiji Kawamura}

\affil[1]{Department of Physics, Nagoya University, Nagoya, Aichi 464-8602, Japan}
\affil[2]{Department of Applied Physics, School of Engineering, The University of Tokyo, 7-3-1 Hongo, Bunkyo-ku, Tokyo 113-8656, Japan}
\affil[3]{Department of Physics, University of Tokyo, Bunkyo-ku, Tokyo 113-0033, Japan}
\affil[4]{Center for Quantum Computing, RIKEN, 2-1 Hirosawa, Wako, Saitama 351-0198, Japan}
\affil[5]{The Kobayashi-Maskawa Institute for the Origin of Particles and the Universe, Nagoya　University, Nagoya, Aichi 464-8602, Japan}

\date{}

\begin{document}
\maketitle

\begin{abstract}
Radiation pressure (RP) noise, one component of quantum noise, can limit the sensitivity of laser interferometric gravitational wave (GW) detectors at lower frequencies. We conceived a possible RP noise cancellation method, using phase flipped ponderomotive-squeezed light (FPSL) incident on free-mass mirrors in interferometers' arms. This possibility is investigated under the constraint that the method is for space-based GW detectors in a broad frequency band lower than 1 Hz without using a long optical cavity. Considering various patterns in a single path small-scale case to generate the FPSL, we proved that no configuration exists in the single path case. 
\end{abstract}


\section{Introduction}
Gravitational wave (GW) detection is one of the most important tools for current and future astronomy. Direct observation of astronomical phenomena using GW is remarkably useful because we can investigate even objects that do not emit electromagnetic waves as long as they move with acceleration. It enables us to observe and investigate such celestial bodies and their phenomena experimentally, which have been explained only theoretically. An opportunity revolutionizing the space research was the first detection of GWs by LIGO \cite{Aasi2015074001} and Virgo \cite{Acernese2015024001}; in 2015, they detected the first black-hole binary merger event \cite{Abbott2016061102}. Two years later, they also detected the first neutron-star binary merger event \cite{Abbott2017161101}. These events provided much information that was hard to get with electromagnetic observations. At present, many countries plan more sophisticated GW detectors \cite{Freise2009085012,Abbott2017044001}. In particular, detection in the low-frequency band is one of the key factors to develop further GW detectors because not only the heavenly bodies but also various cosmological events are targeted at the frequencies. As far as space GW detectors are concerned, there are LISA \cite{Danzmann19971399,Seoane2017786}, DECIGO \cite{Seto2001221103,Kawamura20191845001}, BBO \cite{Ungarelli2005S955}, and so on.

In interferometric GW detectors, laser light is used as a probe to sense GWs. It is because very small distance fluctuations between two free masses, caused by GWs, can be measured as phase fluctuations of the laser light. However, there is a fundamental noise preventing the detection of the GW signals in the laser light: quantum noise \cite{Caves19818}. It comes from quantum fluctuations of the laser light and/or the vacuum itself, and contains two components: shot noise and radiation pressure (RP) noise. The shot noise is a sensor noise; fluctuations in phase quadratures cause fluctuations of photon numbers at a photodetector. On the other hand, the RP noise is displacement noise; it is generated when the laser light with amplitude fluctuations hits free-mass mirrors in the interferometers. The linear spectral density for the shot noise, $\sqrt{S_{\rm shot}}$, and the RP noise, $\sqrt{S_{\rm RP}}$, depend on laser power $P_0$ and the GW frequency $f$, as $\sqrt{S_{\rm shot}} \propto 1/\sqrt{P_0}$ and $\sqrt{S_{\rm RP}} \propto \sqrt{P_0}/f^2$, respectively. From the dependence of the two quantum noises, the GW detector sensitivity is more limited by the RP noise at the lower frequency band especially when the power is increased.

One promising method of reducing the RP noise in a broad frequency band uses a filter cavity \cite{Kimble2001022002}. Ponderomotively-squeezed light generated in the interferometers is injected into the filter cavities, which can modify the level and angle of the squeezed fluctuations in a frequency dependent way. When the modified quantum fluctuations after the filter cavities are detected by the appropriate homodyne detection method, the RP noise can vanish at all frequencies. This method requires filter cavities with a similar size as the arm cavities. However, very long filter cavities could impair the squeezing quality \cite{Iwaguchi20219,Ishikawa202114} due to optical diffraction loss. Therefore, space detectors with long arm cavities such as DECIGO cannot use this method.

In this paper, we consider the possibility of another method: a phase flipped ponderomotive-squeezed light (FPSL) incident on the arms. The FPSL is light in which only the sign of the RP fluctuations is opposite from that of the conventional ponderomotive-squeezed light without the sign of the amplitude and phase fluctuations inverting. Using this method for the ground-based detectors, the RP noise at the photodetector could be completely canceled. In the space-based detectors such as DECIGO, the method of directly injecting the FPSL to the main arm cavities does not work well due to the large optical diffraction loss. Fortunately, a further improvement of sensitivity was suggested by implementing the quantum locking technique with the help of sub-cavities inside the same satellite \cite{Yamada2020126626,Yamada2021127365}. The FPSL can be injected into these short sub-cavities which have negligible optical diffraction loss. Therefore, the FPSL method could be useful for both ground- and space-based detectors.

Some previous works have proposed unique configurations to provide the FPSL \cite{Tsang2010123601,Wimmer2014053836,Moller2017191195}. However, these ideas cannot be applied to DECIGO or similar GW detectors. The techniques in \cite{Tsang2010123601,Wimmer2014053836} require an auxiliary cavity that is as long as the main cavity with a high finesse. The diffraction loss of a cavity as long as DECIGO would greatly impair the squeezing effect of the FPSL. As for the scheme in \cite{Moller2017191195}, lowering the resonant frequency of an atom spin to the GW observation frequency is technically challenging. To find a new configuration providing the FPSL, we searched for a configuration in a single path under the condition that the scale of the system that could produce the FPSL is much smaller than the wavelength corresponding to the GW frequency. The single path is defined as a geometry that a laser path is not divided by a beam splitter (BS). It means that the laser carrier does not contain vacuum fluctuations accompanied with the BS. Optical processes for the configurations, which we covered, are listed as follows:
  \begin{itemize}
    \item the conventional ponderomotive-squeezing caused by the reflection of the light on a free-mass mirror including the possibility of non-normal incident angle,
    \item frequency-independent squeezing with a nonlinear optical medium,
    \item the amplification (or reduction) of the laser amplitude inside an optical cavity, and
    \item all combinations of the above.
  \end{itemize}
In this paper, we mathematically provide an answer to the question, ``Can the phase of radiation pressure fluctuations be flipped in a single path for laser interferometric gravitational waves detectors?''

This paper is organized as follows. In Sec.~\ref{sec:2}, we explain the RP noise-canceling strategy, using the FPSL incident on the interferometer. As a prelude to this description, we discuss the input-output relation of conventional ponderomotive-squeezing. In Sec.~\ref{sec:3}, we define a notation of phase fluctuations coming from the RP, which will be used in Sec.~\ref{sec:4}. Also, we discuss an input-output relation for two other optical processes: frequency-independent squeezing with an optical parametric amplifier (OPA) and changes in the laser effective amplitude in optical cavities or due to non-normal incident angles. In Sec.~\ref{sec:4}, we show step by step that we cannot create the FPSL in the single path case. And finally, in Sec.~\ref{sec:5}, we discuss in more detail the reason why no configuration exists by a case using an optical cavity.

\section{Radiation pressure noise-canceling strategy with phase flipped ponderomotive-squeezed light}
\label{sec:2}
In this section, we explain the core strategy in the paper: the RP noise-canceling strategy using the FPSL. Before considering it, we review the case where conventional vacuum fluctuations are incident from an interferometers' dark port.
  \subsection{Quantum fluctuations and GW signals at the dark port using the conventional vacuum fluctuations as input}
  \label{sec:2-1}
\begin{figure}[htbp]
      \centering
        \includegraphics[clip,width=10.0cm]{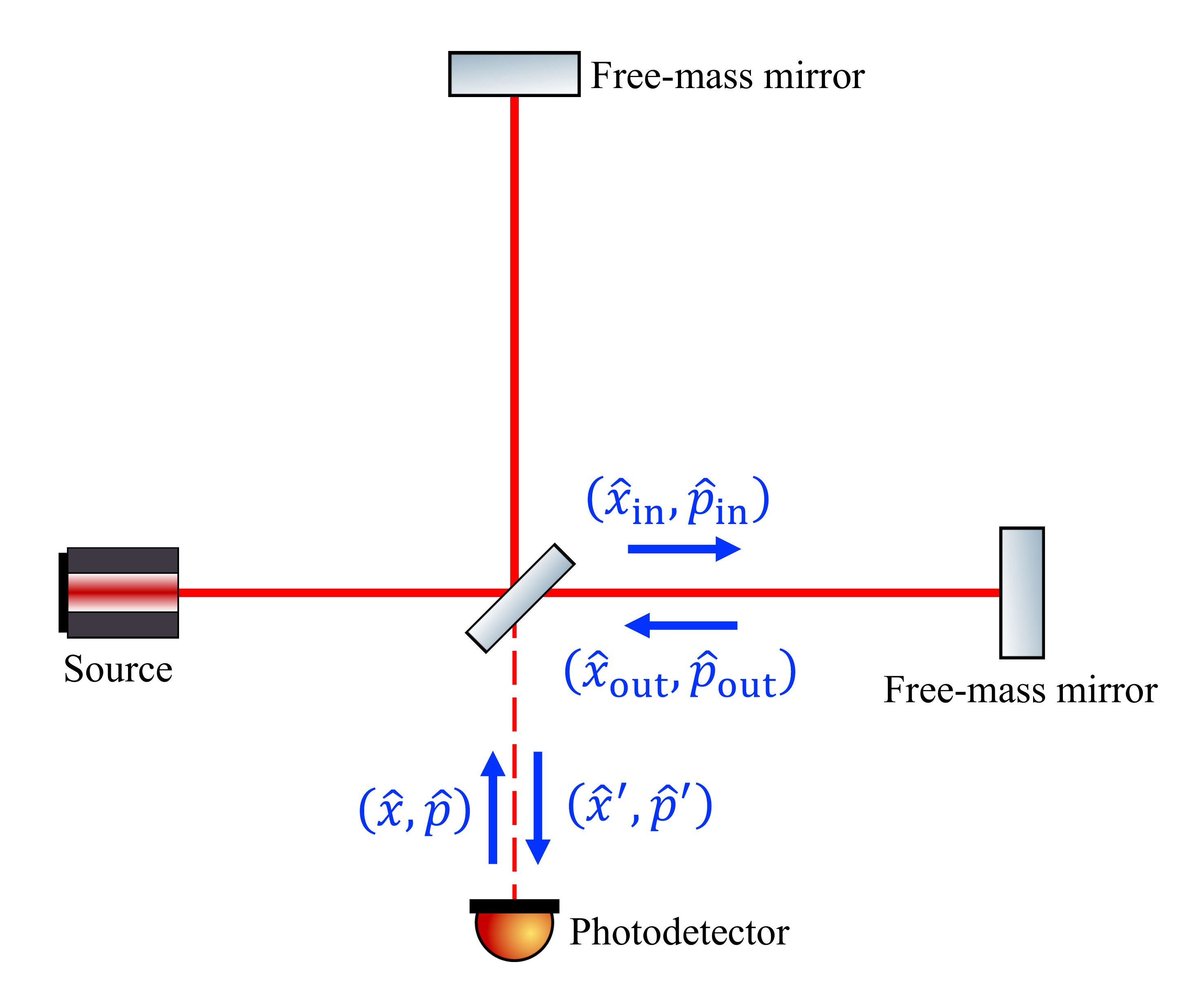}
      \caption{Schematic geometry of a Michelson laser interferometer, showing the notations of quantum fluctuations in some parts of the interferometer. The operators $(\hat{x}, \hat{p})$ and $(\hat{x}', \hat{p}')$ represent the amplitude and phase components of the vacuum fluctuations injected into the interferometer and reflected from the interferometer at the BS along the dark port. The operators $(\hat{x}_{\rm in}, \hat{p}_{\rm in})$ and $(\hat{x}_{\rm out}, \hat{p}_{\rm out})$ show the fluctuations that leave the BS and head for the BS inside the arm of the interferometer.}
      \label{fig:2-0}
    \end{figure}

  We consider a Michelson type laser interferometer, whose fringe at the anti-symmetric port is set as completely dark, as shown in Fig.~\ref{fig:2-0}. In this geometry, the quantum fluctuations at the photodetector are obtained only from the fluctuations incident to the interferometer from the dark port: vacuum fluctuations. We ignore quantum fluctuations inherent in the laser light, and only discuss a relation between the vacuum fluctuations before incident to the interferometer, $(\hat{x}, \hat{p})$, and the fluctuations returned from the interferometer, $(\hat{x}', \hat{p}')$.

  When the laser light enters the interferometer, the vacuum fluctuations from the dark port combine with the laser light at the BS, with opposite signs on the two arms; thus, laser light in each arm of the interferometer contains the vacuum fluctuations $(\hat{x}, \hat{p})$. Here we describe the fluctuations in each interferometer arm as the one-sided electric field operator:
    \begin{equation}
        \hat{E}_+ (z,t) = \sqrt{\frac{2\pi \hbar \omega_0}{\mathcal{A} c}} \mathrm{e}^{- \ai \omega_0 (t - z/c)} \int_0^\infty \left[ \hat{a}_+ \mathrm{e}^{- \ai \Omega (t - z/c)} + \hat{a}_- \mathrm{e}^{+ \ai \Omega (t - z/c)} \right] \frac{\diff \Omega}{2\pi},
      \label{eq:2-1}
    \end{equation}
  where $\omega_0$ is the carrier frequency, $\Omega~(\ll \omega_0)$ is the measurement frequency, and $\mathcal{A}$ is an effective cross-sectional area \cite{Kimble2001022002}. Also $\hat{a}_+, \hat{a}_-$ are annihilation operators and defined using $\omega_0, \Omega$:
    \begin{equation}
        \hat{a}_+ \equiv \hat{a}_{\omega_0 + \Omega},~\hat{a}_- \equiv \hat{a}_{\omega_0 - \Omega}~.
      \label{eq:2-2}
    \end{equation}
  These two annihilation operators $\hat{a}_\pm$ do not represent the amplitude and phase quadratures of the vacuum fluctuations directly; thus, we define operators $\hat{x}_{\rm in}$, $\hat{p}_{\rm in}$ as follows:
    \begin{equation}
        \hat{x}_{\rm in} \equiv \frac{1}{\sqrt{2}} \left( \hat{a}_+ + \hat{a}_-^\dagger \right),~\hat{p}_{\rm in} \equiv - \frac{\ai}{\sqrt{2}} \left( \hat{a}_+ - \hat{a}_-^\dagger \right)~.
      \label{eq:2-3}
    \end{equation}
  Also, we can define the time-varying operators $\hat{x}_{\rm in}(t), \hat{p}_{\rm in}(t)$ by taking the Fourier transform of $\hat{x}_{\rm in}, \hat{p}_{\rm in}$ in Eq.~(\ref{eq:2-3}):
    \begin{align}
      \begin{aligned}
        \hat{x}_{\rm in}(z,t) &= \int_0^\infty \frac{\mathrm{d}\Omega}{2\pi} \left( \hat{x}_{\rm in} \mathrm{e}^{-\mathrm{i} \Omega (t - z/c)} + \hat{x}_{\rm in}^\dagger \mathrm{e}^{\mathrm{i} \Omega (t - z/c)} \right)~, \\
        \hat{p}_{\rm in}(z,t) &= \int_0^\infty \frac{\mathrm{d}\Omega}{2\pi} \left( \hat{p}_{\rm in} \mathrm{e}^{-\mathrm{i} \Omega (t - z/c)} + \hat{p}_{\rm in}^\dagger \mathrm{e}^{\mathrm{i} \Omega (t - z/c)} \right)~.
      \end{aligned}
      \label{eq:2-3-2}
    \end{align}
  $\hat{x}_{\rm in}(z,t),~\hat{p}_{\rm in}(z,t)$ in Eq.~(\ref{eq:2-3-2}) are the quadrature-phase amplitudes of time, and $\hat{x}_{\rm in},~\hat{p}_{\rm in}$ in Eq.~(\ref{eq:2-3}) are the Fourier components of $\hat{x}_{\rm in}(z,t),~\hat{p}_{\rm in}(z,t)$. We will use the Fourier components $\hat{x}_{\rm in},~\hat{p}_{\rm in}$ so that all calculations are discussed in the frequency domain in this paper. The two fluctuations' operators $(\hat{x}_{\rm in},~\hat{p}_{\rm in})$ correspond to the operators in the arm of the interferometer in Fig.~\ref{fig:2-0}.

\begin{figure}[htbp]
      \centering
        \includegraphics[clip,width=10.0cm]{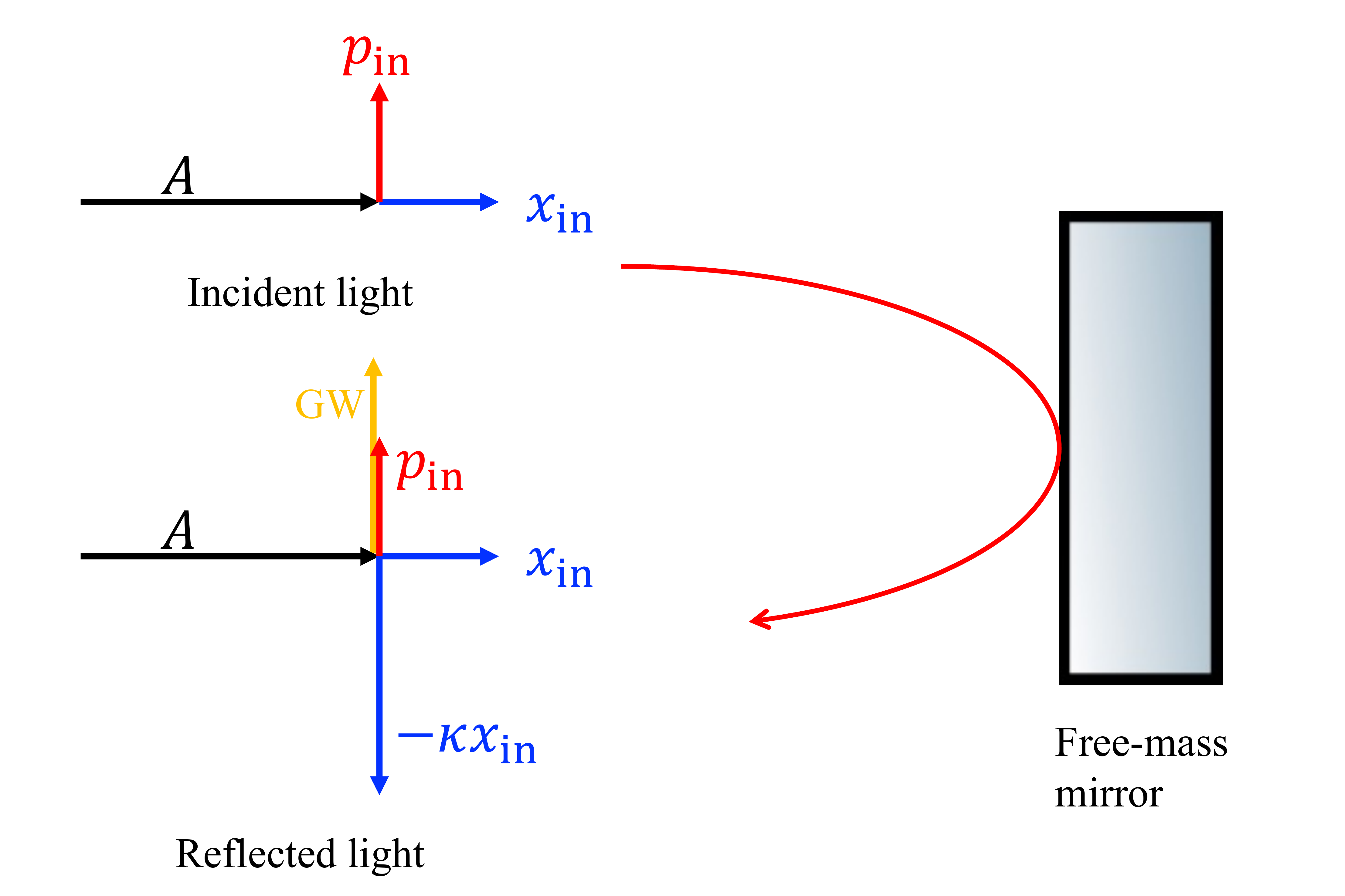}
      \caption{Input-output relation for the optomechanical process at the free-mass mirror in either arm of the interferometer. The black arrow $(A)$ means the carrier amplitude. The blue $(x_{\rm in})$ and the red $(p_{\rm in})$ arrows in the upper left part of the figure represent the amplitude and phase fluctuations, respectively. The RP fluctuations $-\kappa x_{\rm in}$  are caused by the amplitude fluctuations $x_{\rm in}$; thus, we depict the two vectors using the same color (blue) as for $x_{\rm in}$. The relative phase between the carrier and the fluctuations in the other arm of the interferometer is opposite because the fluctuations coming from the dark port are differentially incident to each arm.}
      \label{fig:2-1}
    \end{figure}

  We consider the laser light with two components of fluctuations, $(\hat{x}_{\rm out}, \hat{p}_{\rm out})$, after being reflected from an ideal free-mass mirror whose reflectivity is perfect \cite{Caves19818}. Because RP from the time-varying component of the laser power causes test mass motion which leads correlated additional phase fluctuations, an input-output relation between $(\hat{x}_{\rm in},\hat{p}_{\rm in})$ and $(\hat{x}_{\rm out},\hat{p}_{\rm out})$ contains the following off-diagonal component:
    \begin{equation}
        \begin{pmatrix}
          \hat{x}_{\rm out} \\ \hat{p}_{\rm out}
        \end{pmatrix} =
        \begin{pmatrix}
          1 & 0 \\ -\kappa & 1
        \end{pmatrix}
        \begin{pmatrix}
          \hat{x}_{\rm in} \\ \hat{p}_{\rm in}
        \end{pmatrix}~.
      \label{eq:2-4}
    \end{equation}
  This relation is shown in Fig.~\ref{fig:2-1}, using a phasor diagram. The coupling constant $\kappa$ is expressed as
    \begin{equation}
        \kappa \equiv 2 A k_0 \cdot \frac{1}{M\Omega^2} \cdot \frac{2\hbar\omega_0 A}{c}~,
      \label{eq:2-5}
    \end{equation}
  where $A$ is the carrier amplitude of the laser light, $k_0$ is the wave number, and $M$ is the mass of the mirror. Equation (\ref{eq:2-5}) consists of three factors: the last factor denotes the RP forces caused by fluctuating laser power, the second denotes the transfer function from applied force to test mass displacement, and the first converts the angular rotation of the incident carrier to a phase fluctuation phasor. In the following, we refer to the phase fluctuations $-\kappa \hat{x}_{\rm in}$ as RP fluctuations to distinguish them from the vacuum phase fluctuations.

  At the BS, either quadrature of the quantum fluctuations incident on the BS are related to the same quadrature of the quantum fluctuations after the BS. Thus from Eq.~(\ref{eq:2-4}), the input-output relation between the vacuum fluctuations incident on the dark port of the interferometer and the fluctuations returning from the interferometer can be written as:
    \begin{equation}
        \begin{pmatrix}
          \hat{x}' \\ \hat{p}'
        \end{pmatrix} =
        \begin{pmatrix}
          1 & 0 \\ -\kappa & 1
        \end{pmatrix}
        \begin{pmatrix}
          \hat{x} \\ \hat{p}
        \end{pmatrix}~.
      \label{eq:2-4-2}
    \end{equation}
  Comparing $(\hat{x},\hat{p})$ and $(\hat{x}', \hat{p}')$, the vacuum fluctuations are squeezed frequency-dependently. This unique frequency dependence of the squeezing is called as the ponderomotive-squeezing. The squeezing level and angle of the ponderomotive-squeezing are obtained from the phase quadratures of fluctuations which occur from the amplitude quadratures, $-\kappa \hat{x}$.

  We consider the noise and GW components in the vacuum fluctuations returned from the interferometer $(\hat{x}', \hat{p}')$. As for the noise components, it is described with $(\hat{x}, \hat{p})$ from Eq.~(\ref{eq:2-4-2}):
    \begin{equation}
        \hat{p}' = \hat{p} - \kappa \hat{x}~.
      \label{eq:2-6}
    \end{equation}
  GW signals $\hat{p}_\mathrm{GW}$ are also added, whose amplitude is proportional to the arm length $L$ when the GW wavelength is much longer than $L$. The sign of $\hat{p}_\mathrm{GW}$ is opposite for the two perpendicular arms, thus it remains in $\hat{p}'$ at the dark port. The phase fluctuations $\hat{p}'$ are expressed by adding $\hat{p}_{\rm GW}$ to Eq.(\ref{eq:2-6}):
    \begin{equation}
        \hat{p}' = \hat{p} - \kappa \hat{x} + \hat{p}_\mathrm{GW}~.
      \label{eq:2-7}
    \end{equation}
  It obviously shows that the phase fluctuations $\hat{p}$ and the RP fluctuations that generated from the amplitude fluctuations $-\kappa \hat{x} \propto f^{-2}$ interferes with detecting the GW signals $(\hat{p}_\mathrm{GW})$.

  \subsection{Quantum fluctuations and GW signals at the dark port using phase flipped ponderomotive-squeezed fluctuations as input}
  \label{sec:2-2}
\begin{figure}[htbp]
      \centering
        \includegraphics[clip,width=10.0cm]{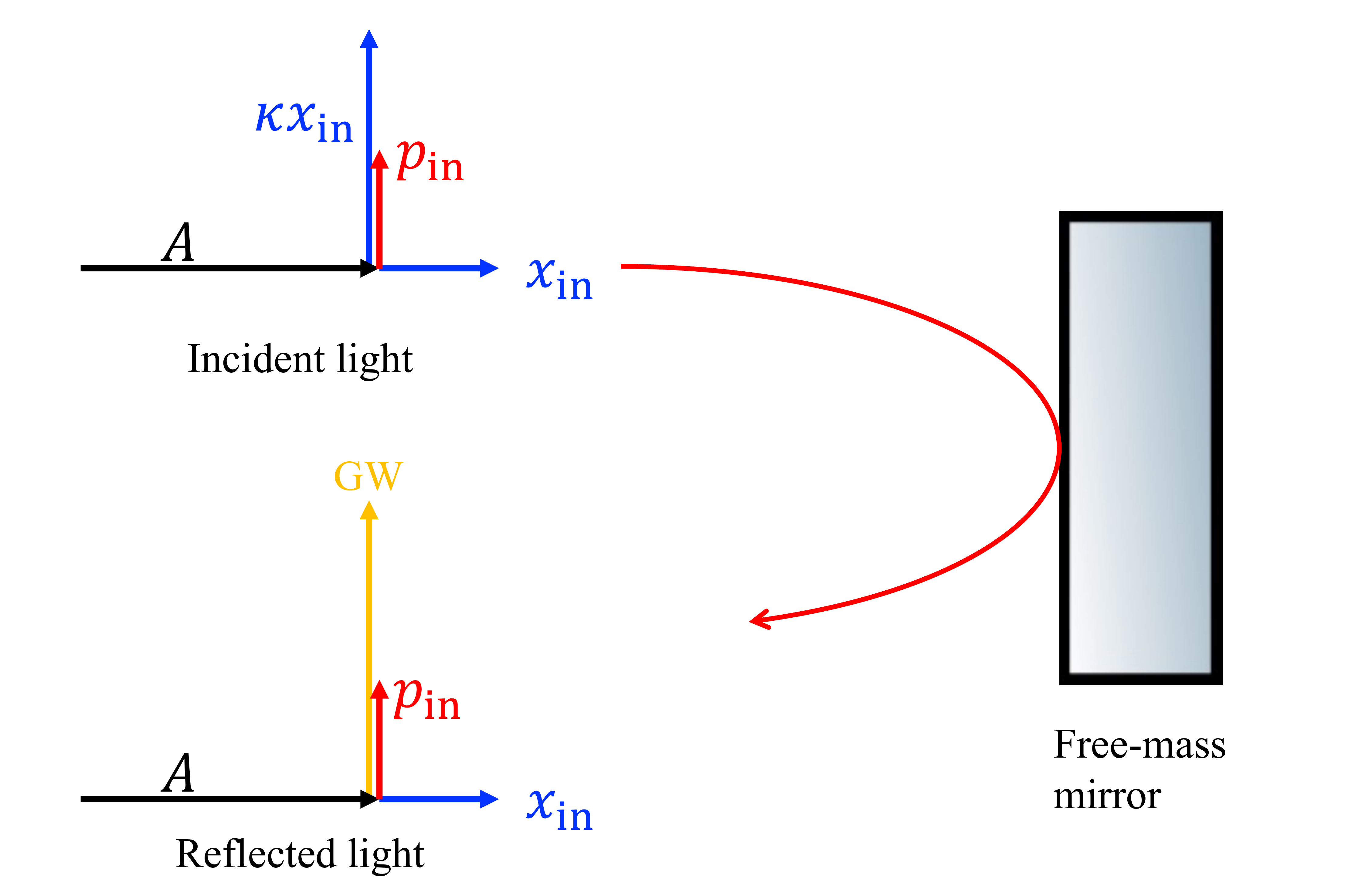}
      \caption{Phasor Diagram for the reflected light from the interferometer's arms with FPSL. We sketch the case in which the squeezing level and angle are the same as those generated from the mirror ($\tilde{\kappa} = \kappa$). Phase fluctuations previously created (blue arrow in the incident light) are compensated with the newly-added RP fluctuations at the mirror placed in the arms.  Only the vacuum phase fluctuations resulting in the shot noise and the GW signals remain.}
      \label{fig:2-3}
    \end{figure}

  We consider the case in which the FPSL hits the mirror placed at the end of the interferometers' arms. We can express the incident vacuum fluctuations in the dark port as $(\hat{x}, \hat{p} + \tilde{\kappa} \hat{x})$. This assumption comes from the conventional ponderomotive-squeezed fluctuations, which can be written as $(\hat{x}, \hat{p} - \kappa \hat{x})$ in Eq.~(\ref{eq:2-4-2}). Figure \ref{fig:2-3} represents an input-output relation between before and after hitting the interferometer's arms, and it shows that the phase flipped RP fluctuations prepared in advance $\tilde{\kappa} \hat{x}$ are counteracted to a newly-added RP fluctuations $-\kappa \hat{x}$ caused from the amplitude fluctuations shaking the mirrors. We can generally express the input-output relation for the vacuum fluctuations in the same way as Eq.~(\ref{eq:2-4-2})
    \begin{align}
        \begin{pmatrix}
          \hat{x}' \\ \hat{p}'
        \end{pmatrix} =
        \begin{pmatrix}
          1 & 0 \\ -\kappa & 1
        \end{pmatrix}
        \begin{pmatrix}
          \hat{x} \\ \hat{p} + \tilde{\kappa} \hat{x}
        \end{pmatrix}~.
      \label{eq:2-8}
    \end{align}
  Including the GW signals $\hat{p}_\mathrm{GW}$, the reflected phase quadratures $\hat{p}'$ are given by
    \begin{equation}
        \hat{p}' =  \hat{p} - \left( \kappa - \tilde{\kappa} \right) \hat{x} + \hat{p}_\mathrm{GW}~.
      \label{eq:2-9}
    \end{equation}

  If the squeezing level and angle of the incident light are properly adjusted ($\tilde{\kappa} = \kappa$), the two RP fluctuations cancel each other. Note that the mirror motion itself is not canceled by the FPSL. Only the newly-added RP fluctuations from the mirror motion are counteracted to the phase flipped RP fluctuations. In Eq.~(\ref{eq:2-9}), the coefficient of the third term, $\hat{p}_\mathrm{GW}$, is not changed from Eq.~(\ref{eq:2-7}); thus, the FPSL can sense GWs in the same way as the ordinary laser light. On the sensitivity curve, the shot noise is not changed, but the RP noise becomes ``zero''. Therefore, the curve depends only on the shot noise at all frequencies.

\section{Notation for optical interaction}
In this section, we discuss input-output relations of some optical processes, and we define a notation of phase fluctuations coming from the RP, for the discussion in Sec.~\ref{sec:4}.

\label{sec:3}
  \subsection{Squeezing with a nonlinear optical medium}
\begin{figure}[htbp]
      \centering
        \includegraphics[clip,width=10.0cm]{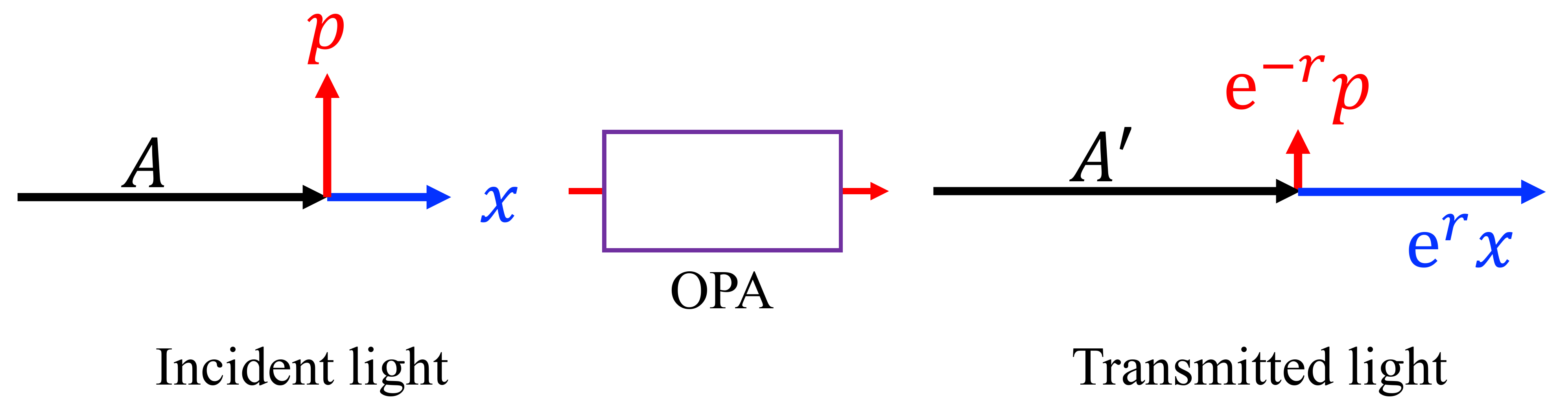}
      \caption{Input-output relation for squeezing process using an optical parametric amplifier (OPA). We assume $r = 0.69~(\mathrm{e}^r \approx 2)$. In this case, fluctuations are squeezed vertically. The laser carrier $A$ is also amplified by the OPA.}
      \label{fig:2-2}
    \end{figure}

  A nonlinear optical process can squeeze the quantum fluctuations. For example, an optical parametric amplifier (OPA) causes frequency-independent squeezing \cite{Lugiato19835}. The squeezing operator $\hat{S}(r,\phi)$ is written in \cite{Caves19853068,Schumaker19853093}:
    \begin{equation}
        \hat{S}(r,\phi) \equiv \exp \left[ r \left( \hat{a}_+ \hat{a}_- \mathrm{e}^{-2\ai\phi} - \hat{a}_+^\dagger \hat{a}_-^\dagger \mathrm{e}^{2\ai\phi} \right) \right]~,
      \label{eq:2-14}
    \end{equation}
  where $r$ is the squeezing level, and $\phi$ is the squeezing angle. Using this equation, we can obtain the input-output relation for the two quadratures:
    \begin{equation}
        \begin{pmatrix}
          \hat{x}_{\rm sv} \\ \hat{p}_{\rm sv}
        \end{pmatrix} =
        \begin{pmatrix}
          \cosh r + \sinh r \cos(2\phi) & \sinh r \sin(2\phi) \\ \sinh r \sin(2\phi) & \cosh r - \sinh r \cos(2\phi)
        \end{pmatrix}
        \begin{pmatrix}
          \hat{x} \\ \hat{p}
        \end{pmatrix}~.
      \label{eq:2-15}
    \end{equation}
  If we assume $\phi = 0$ for simplicity, Eq.~(\ref{eq:2-15}) is replaced by
    \begin{equation}
        \begin{pmatrix}
          \hat{x}_{\rm sv} \\ \hat{p}_{\rm sv}
        \end{pmatrix} =
        \begin{pmatrix}
          \mathrm{e}^r & 0 \\ 0 & \mathrm{e}^{-r}
        \end{pmatrix}
        \begin{pmatrix}
          \hat{x} \\ \hat{p}
        \end{pmatrix}~.
      \label{eq:2-16}
    \end{equation}
  It obviously shows that the phase quadratures are decreased, while the amplitude quadratures are increased when $r > 0$. Figure \ref{fig:2-2} shows its schematic diagram.

  \subsection{Effective laser amplitude}
  \label{sec:2-3}
  In this subsection, we discuss how some optical interactions can be described using an effective laser amplitude, which includes effects of the amplitude fluctuations' changes in the optical interactions. We consider two cases as examples: the laser light incident to the free-mass mirror with a non-normal angle, and the light incident on an optical cavity.

  Considering the first case, we assume an incident angle to the mirror $\theta$. The time-varying RP on the mirror weakens by $\cos\theta$ because only the normal component of the photon momentum is altered by the reflection. A second factor $\cos\theta$ is introduced because the mirror motion results in a smaller pathlength change along the laser path. As a result, Eq.~(\ref{eq:2-5}) is modified from the vertically incident case as
    \begin{equation}
        \kappa = 2 A k_0 \cos\theta \cdot \frac{1}{M\Omega^2} \cdot \frac{2\hbar\omega_0 A}{c} \cos\theta~.
      \label{eq:2-17}
    \end{equation}
  Therefore, we can substitute an effective amplitude $A_{\rm eff}^{\rm diag}$ for the laser amplitude $A$ in Eq.~(\ref{eq:2-5}), the normal incidence case ($\theta = 0$), with:
    \begin{equation}
        A_{\rm eff}^{\rm diag} = A \cos\theta~.
      \label{eq:2-18}
    \end{equation}

  In the case of a cavity case, we assume that the reflectivity of the end mirror is set as unity while that of the input mirror is $r_{\rm in} = r~(\neq 1)$. Otherwise, the laser path can be divided at the cavity if the end mirror has the reflectivity $r_{\rm end}~(\neq 1)$, which violates our assumption of a single path. We allow for the case of multiple mirrors in the cavity provided all except the input mirror have reflectivity equal to 1. Furthermore, we assume that the cavity is in perfect resonance for simplicity. Ignoring any optical loss, we can rewrite the $A$ in a similar manner as Eq.~(\ref{eq:2-18}) using the input mirror reflectivity $r$:
    \begin{equation}
        A_{\rm eff}^{\rm cavity} = \left( \frac{1 + r}{t} \right)^2 A~.
      \label{eq:2-19}
    \end{equation}
  It is composed of three effects. One of them is an amplification effect of laser amplitude inside the cavity. The second is the amplification of the amplitude quadrature fluctuations inside the cavity. And the third is the number of times the carrier is phase-shifted by the mirror displacement. In the high-finesse case, where the reflectivity approaches unity, $A_{\rm eff}^{\rm cavity}$ is approximated to $(\mathcal{F}/2) A$.

    \subsection{Difference of radiation pressure fluctuations from the laser hitting on either side of the mirror surface}
  \label{sec:3-1}
  In this subsection, we consider the RP fluctuations $-\kappa \hat{x}$ from the point of view of which surfaces of the free-mass mirror the laser light hits. Then we will describe a simple notation of the RP fluctuations. We write coefficients corresponding to the two operators $\hat{x}, \hat{p}$ as $X, P$. Here and after this section, we consider a single measurement frequency $\Omega$.

  As for the carrier amplitude in Eq.~(\ref{eq:2-5}), $\kappa$ is proportional to $A^2$. The carrier amplitude $A$ in the last factor of the right-hand side of Eq.~(\ref{eq:2-5}) is responsible for causing RP motion, while the $A$ in the first factor is needed only to read out the mirror motion. We should distinguish the roles of the two carrier amplitudes.

  We consider the RP fluctuations in the case where the beam which senses the RP motion is the same as the one causing it (Fig.~\ref{fig:2-1}). The RP fluctuations in Fig.~\ref{fig:2-1} can be written in a simple form using the carrier amplitude $A_{\rm front}$:
    \begin{equation}
        P_{\rm front, front} \propto - \frac{1}{M} {A_{\rm front}}^2 X_{\rm front}~.
      \label{eq:2-10}
    \end{equation}

  We also derive the RP fluctuations in the case where the mirror motion is generated from the laser hitting on the front, and then the motion is read out from the laser hitting on the back of the mirror. Although the motion is also picked up from the laser hitting on the front, we concentrate on the case the motion picked up from the opposite side of the mirror as we have obtained the notation in that case: Eq.~(\ref{eq:2-10}). Comparing this case with the case discussed before, $A$ and its sign in the first factor are different. It is because, in this case, the laser light recording the mirror motion senses as if the mirror was pulled when the laser power pushed the mirror on the front. As a result, the form of the RP fluctuations in this case slightly differs from Eq.~(\ref{eq:2-10}):
    \begin{equation}
        P_{\rm front, back} \propto - \frac{1}{M} \left( - A_{\rm back} \right) \cdot A_{\rm front} X_{\rm front}~.
      \label{eq:2-11}
    \end{equation}

  We consider the amplitude operator $\hat{x}$, after the $i$-th optical processes, as $\hat{x}_i$. We also consider the RP fluctuations $\hat{p}_{\rm RP} (= - \kappa \hat{x})$ the mirror motion is generated at the $i$-th optical process and is recorded at the $j$-th optical process as $(\hat{p}_{\rm RP})_{i,j}$. Also, we use these coefficients as $X_i, P_{i,j}$. Combining the two cases mentioned above, we generally define a simple form of the RP fluctuations as
    \begin{equation}
        P_{i, j} \equiv - \frac{1}{M} \delta_j A_j \cdot \delta_i A_i X_i~.
      \label{eq:2-12}
    \end{equation}
  Here $i,j$ are the indices of the optical processes, and $\delta_i$ represents how the light interacts with the mirror at the $i$-th optical process:
    \begin{align}
        \delta_i = \left\{
        \begin{array}{cl}
          +1 & (\text{Light incident on the mirror from the front})~, \\
          -1 & (\text{Light incident on the mirror from the back})~, \\
          0 & (\text{Light does not hit the mirror})~.
        \end{array} \right.
      \label{eq:2-13}
    \end{align}
  Note that $P_{i,j}$ takes a constant factor $2k_0 / \Omega^2 \cdot 2 \hbar \omega_0 / c$ from Eq.~(\ref{eq:2-10}). The sign of the factor is positive; thus, the sign of $P_{i, j}$ follows the sign in the right-hand side of Eq.~(\ref{eq:2-12}). We will use this notation of the RP fluctuations in Sec.~\ref{sec:4} and Sec.~\ref{sec:5}.

\section{Single path}
\label{sec:4}
In this section, we will show that the RP fluctuations in a single path cannot create the FPSL, starting from a few special cases and finally concluding as a general case. In the following, we discuss cases divided by the number of mirrors, optical processes, and squeezer using a nonlinear optical medium.

  \subsection{Case: 1 mirror, $n$ optical processes, and 1 squeezer}
  \label{sec:4-1}
  To begin, we explain a basic strategy of proving that any configuration cannot create the FPSL. First, we consider the amplitude fluctuations $X_i$ after the $i$-th optical process ($i \leqq n$). Second, using $X_i$ from step 1, we calculate the total RP fluctuations, $P_{\rm total}$, via the net mirror displacement induced by the fluctuating RP, $\Delta x_{\rm total}$. Finally, we calculate $\mathrm{sgn} (X_n P_{\rm total})$. Using the notation of the phase flipped ponderomotive-squeezed fluctuations in Sec.~\ref{sec:2-2}, the sign of the product could be written as
      \begin{equation}
          \mathrm{sgn} (X_n P_{\rm total}) = +1~.
        \label{eq:3-1}
      \end{equation}
  We will consider the feasibility of the FPSL, that is, the sign of the product $\mathrm{sgn} (X_n P_{\rm total})$.

\begin{figure}[htbp]
    \centering
      \includegraphics[clip,width=12.0cm]{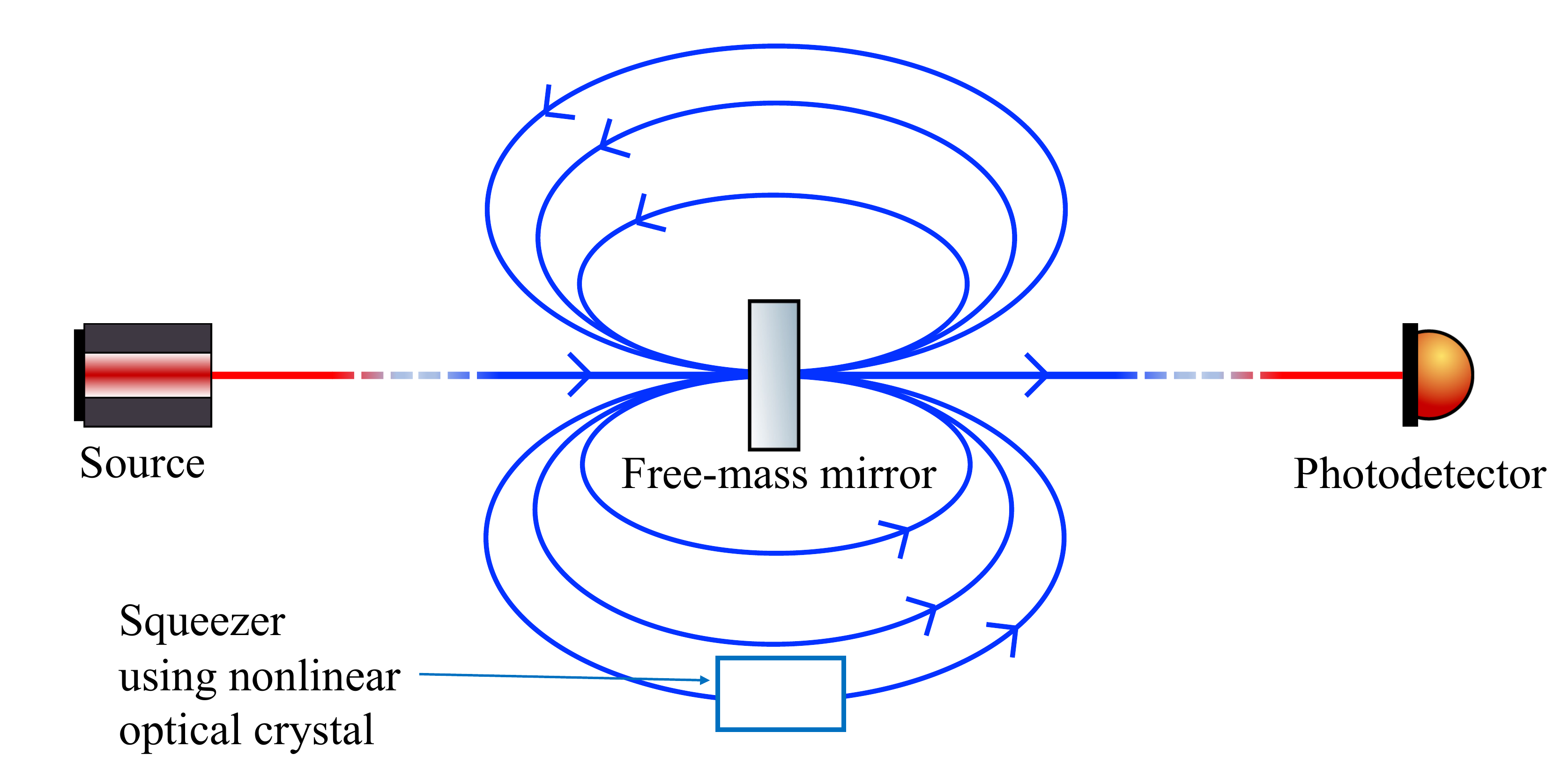}
    \caption{Schematic geometry of the case: 1 mirror, $n$ processes, and 1 squeezer. In the $n_1$-th process, the laser passes through the squeezer, and all the other processes are hitting the mirror from its front or back. Although we illustrate that the laser hits the front and back of the mirror alternately, the mathematical formulation allows any individual reflection to occur from either side. The blue paths hitting the free mass represent the generally changed (including not changed) carrier effective amplitude by optical processes.}
    \label{fig:3-1}
  \end{figure}

  In this subsection, we discuss the following case: 1 mirror, $n$ optical processes, including 1 squeezer. We assume that the laser quantum fluctuations are squeezed elliptically at the $n_1$-th $(n_1 \leqq n)$ optical process, as shown in Fig.~\ref{fig:3-1}. We consider the amplitude fluctuations at the $i$-th optical process $X_i$ separately before and after the crystal squeezer. Before the squeezer $(1 \leqq i < n_1)$, none of the optical processes (reflections from the mirror) change the amplitude fluctuations. Passing through the squeezer, the amplitude fluctuations are changed by a factor of $\mathrm{e}^r$. Accordingly, the amplitude fluctuations' coefficients $X_i$ are:
    \begin{align}
        X_i = \left\{
          \begin{array}{cc}
            X & (1 \leqq i < n_1)~, \\
            \mathrm{e}^r X & (n_1 \leqq i \leqq n)~.
          \end{array}
        \right.
      \label{eq:3-2}
    \end{align}
  Note that a relative phase difference between the two sidebands $\hat{a}_\pm$ and the carrier amplitude $A_i$ causes $X_i$ to deviate from 1. It leads to a symbol $X$ taking a value from -1 to 1. $X$ is independent of the index $i$ because we assume that the scale of configuration is much smaller than the wavelength of the sidebands.

  Next, the mirror displacement induced from the $i$-th optical process (laser reflection from the mirror) is derived by using $X_i$ and $A_i$ in the same discussion as Sec.~\ref{sec:3-1}
    \begin{equation}
        \Delta x_i = - \frac{1}{M} \delta_i X_i A_i~,~~(i = 1,2,\cdots,n)~.
      \label{eq:3-3}
    \end{equation}
  The minus sign comes from the transfer function from the applied force to the displacement: $-1/M\Omega^2$. Since a displacement change is generated whenever the laser light hits the mirror, and since $\delta_i$ in Eq.~(\ref{eq:2-13}) denotes whether the light hits or not as well as which surface of the mirror the light hits, the net displacement is given by
    \begin{equation}
        \Delta x_{\rm total} = \sum_{i=1}^n \Delta x_i = - \frac{1}{M} \sum_{i=1}^n \delta_i X_i A_i~.
      \label{eq:3-4}
    \end{equation}

  Finally, on each reflection the motion of the mirror (Eq.~(\ref{eq:3-4})) causes a phase shift of the classical amplitude to produce a phase fluctuation. We can obtain the net RP fluctuations at the $j$-th optical process according to Eq.~(\ref{eq:2-12}):
    \begin{equation}
        P_j = + \delta_j A_j \Delta x_{\rm total}~,~~(j = 1,2,\cdots,n)~.
      \label{eq:3-5}
    \end{equation}
  The plus sign represents that the reflected laser phase fluctuations $P_j$ are delayed by extending the laser path $(\Delta x_n > 0)$. Considering the total RP fluctuations at a photodetector $(i = n)$, it is incomplete to simply sum up all the RP fluctuations $P_j$ in Eq.~(\ref{eq:3-5}). $P_j~(j < n_1)$ passes through the squeezer and is changed by the factor of $\mathrm{e}^{-r}$. On the other hand, $P_j$ generated after the squeezer $(j > n_1)$ has no change. For this reason, the $j$-th component in Eq.~(\ref{eq:3-5}) should be corrected by Eq.~(\ref{eq:3-2}), and the coefficient for the total RP fluctuations $P_{\rm total}$ is given by
    \begin{equation}
        P_{\rm total} = \sum_{j = 1}^n \frac{X_j}{X_n} P_j = - \frac{1}{M} \cdot \frac{1}{X_n} \left( \sum_{i = 1}^n \delta_i X_i A_i \right)^2~.
      \label{eq:3-7}
    \end{equation}
  We examine the sign of $X_n P_{\rm total}$ as follows:
    \begin{equation}
        \mathrm{sgn} (X_n P_{\rm total}) = \mathrm{sgn} \left[ - \frac{1}{M} \left( \sum_{i = 1}^n \delta_i X_i A_i \right)^2 \right] \neq +1~.
      \label{eq:3-8}
    \end{equation}
  It shows that the RP fluctuations cannot be inverted from the conventional one in this configuration.

  \subsection{Expansion from 1 squeezer to multiple squeezers}
  \label{sec:3-2}
\begin{figure}[htbp]
      \centering
        \includegraphics[clip,width=12.0cm]{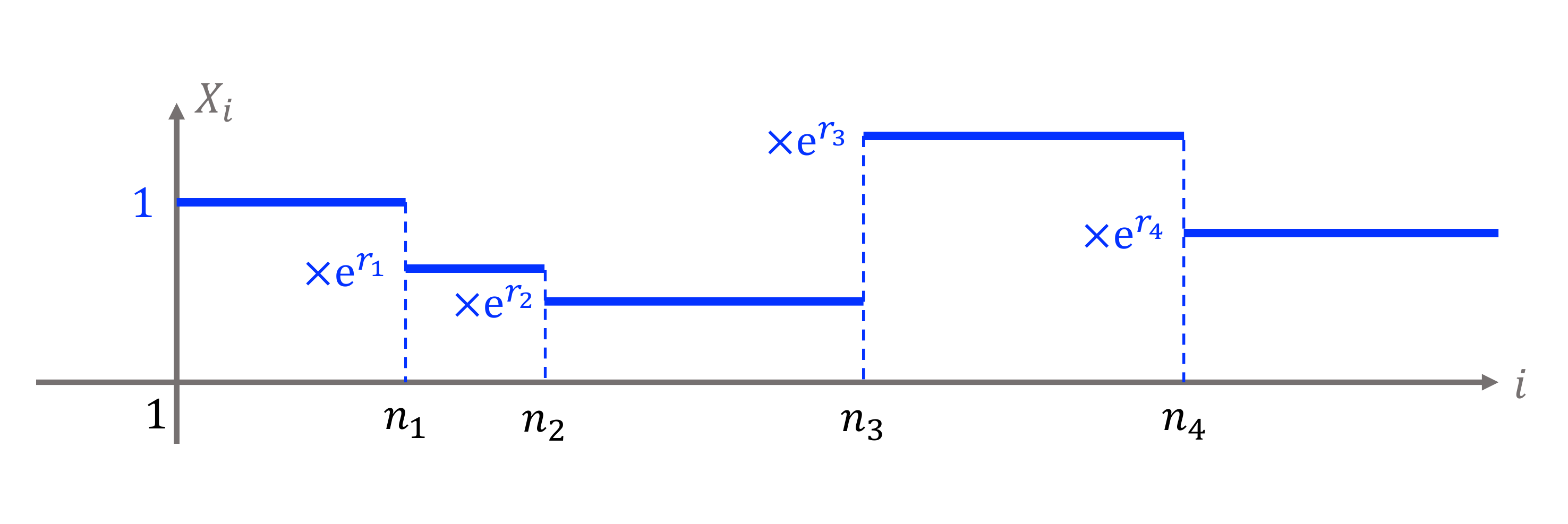}
      \caption{Graph of the coefficient $X_i$ in the multiple squeezer case. The amplitude of $X_i$ is suddenly changed every time the laser light goes through a squeezer, and an additional squeezing factor $\mathrm{e}^{r}$ is multiplied.}
      \label{fig:3-2}
    \end{figure}

  We expand the discussion in the previous subsection to the multiple squeezers, i.e., from the 1-time squeezing case to the $l~(\leqq n)$-times case. The amplitude fluctuations are changed whenever the laser light passes through a squeezer, and a new factor $\mathrm{e}^r$ is multiplied to the previous degree of the amplitude fluctuations. Figure~\ref{fig:3-2} shows that the graph of $X_i$ as a function of $i$ depicts the additional multiplication as a sudden change of the amplitude of $X_i$. Therefore, the modified coefficient $X_i$ including multiple squeezing effects is written as follows:
    \begin{align}
        X_i = \left\{
          \begin{array}{cl}
            X & (1 \leqq i < n_1)~, \\
            \mathrm{e}^{r_1} X & (n_1 \leqq i < n_2)~, \\
            \mathrm{e}^{r_1} \mathrm{e}^{r_2} X & (n_2 \leqq i < n_3)~, \\
            \vdots & \\
            {\displaystyle \left( \prod_{k = 1}^{n_l} \mathrm{e}^{r_k} \right) X} & (n_l \leqq i \leqq n)~.
          \end{array}
        \right.
      \label{eq:3-9}
    \end{align}
  Since we can have the same discussion as Eq.~(\ref{eq:3-2}), only by changing $X_i$ from Eq.~(\ref{eq:3-2}) to Eq.~(\ref{eq:3-9}), the sign of the product, $\mathrm{sgn} (X_n P_{\rm total})$, is the same form as Eq.~(\ref{eq:3-8}). Therefore, we find that it leads to the same conclusion as before even if the squeezing occurs multiple times.

  \subsection{General case: further expansion from 1 mirror to multiple mirrors}
  \label{sec:3-3}
\begin{figure}[thbp]
      \centering
        \includegraphics[clip,width=12.0cm]{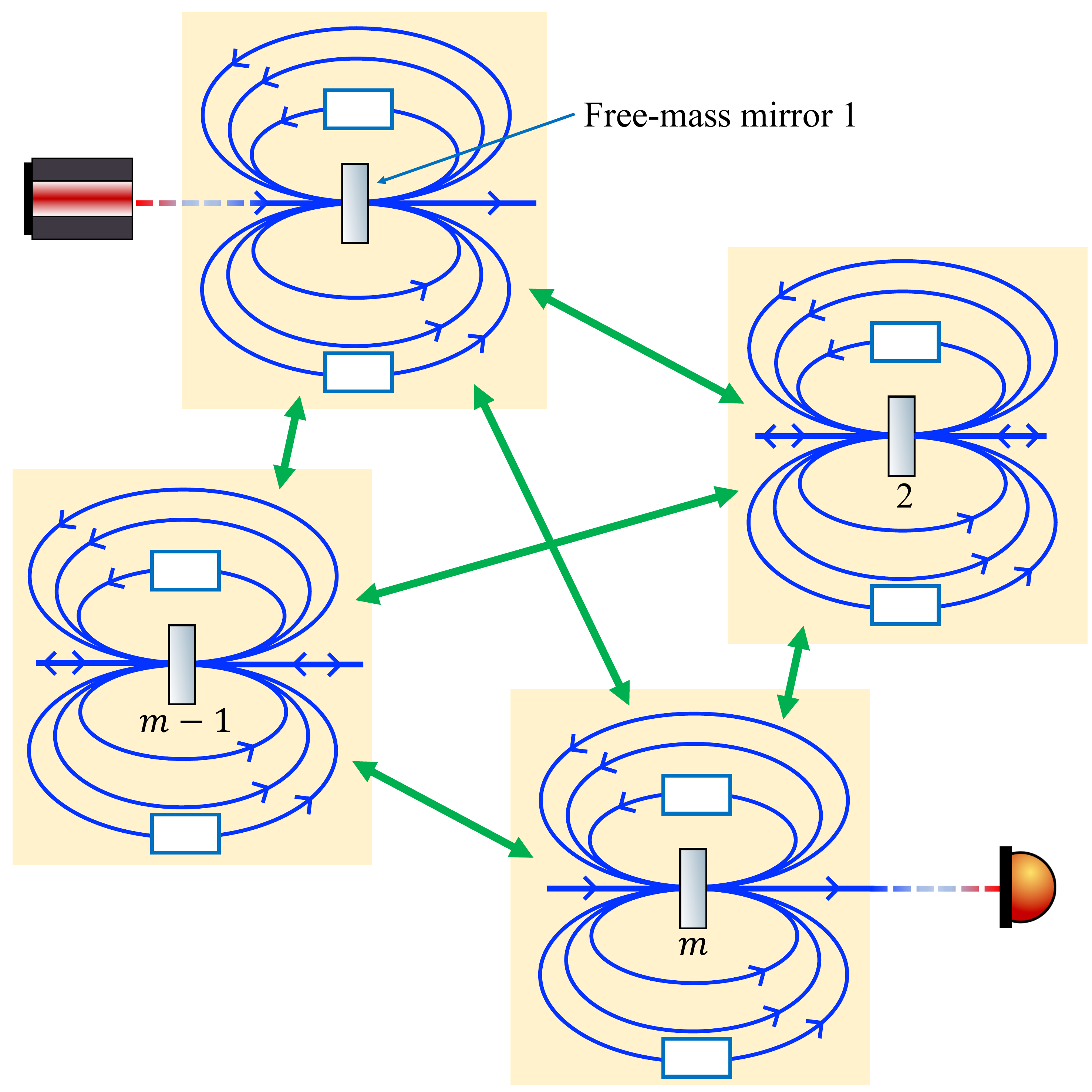}
      \caption{Connection among each mirror using the laser beam. The laser light can hit each mirror with an arbitrary coupling coefficient in an arbitrary order. Also, the laser light can pass each squeezer with an arbitrary squeezing coefficient in an arbitrary order. Although the drawing shows the alternately-hitting cases as Fig.~\ref{fig:3-1}, the discussion in Sec.~\ref{sec:3-3} can treat all the cases in a single path.}
      \label{fig:3-3}
    \end{figure}

  We also expand the discussion to multiple ($m$) mirrors: this geometry, including a cavity case, is completely general in the single path. A schematic picture is shown in Fig.~\ref{fig:3-3}. The net mirror motion is determined by the two conditions: whether the laser light hits the certain mirror or not and how the light hits the mirror, no matter in what order they occur. Moreover, the motion of each mirror uniquely differs from mirror to mirror. It leads to the modification of the interaction factor $\delta_i$. It should be different for each mirror: $\delta_{i,\alpha}~(\alpha = 1,2,\cdots m)$. We redefine the notation $\delta_{i,\alpha}$ as
    \begin{align}
        \delta_{i,\alpha} = \left\{
        \begin{array}{cl}
          +1 & (\text{Incident light pushes mirror $\alpha$ from the front})~, \\
          -1 & (\text{Incident light pushes mirror $\alpha$ from the back})~, \\
          0 & (\text{Light does not hit mirror $\alpha$})~.
        \end{array} \right.
      \label{eq:3-10}
    \end{align}
  As for the subtotal RP fluctuations of each mirror $P_{{\rm total},\alpha}$, the previous discussion in Sec. \ref{sec:3-2} is individually applied to each mirror in this case. Setting the mass of mirror $\alpha$ as $M_\alpha$, the total RP fluctuations can be obtained by the sum of all the RP fluctuations generated from each mirror.
    \begin{equation}
        P_{\rm total} = \sum_{\alpha = 1}^m P_{{\rm total},\alpha} = - \frac{1}{X_n} \sum_{\alpha = 1}^m \frac{1}{M_\alpha} \left( \sum_{i = 1}^n \delta_{i,\alpha} X_i A_i \right)^2~.
      \label{eq:3-11}
    \end{equation}
  Note that the reason why we take a linear sum, but not take the quadrature sum, is that the RP fluctuations of each mirror come from the same amplitude fluctuations of the laser light. From $X_n$ in Eq.~(\ref{eq:3-9}) and $P_n$ in Eq.~(\ref{eq:3-11}), we can see that the ponderomotive-squeezing cannot be inverted in polarity with any configurations in the above scope
    \begin{equation}
        \mathrm{sgn} (X_n P_{\rm total}) = \mathrm{sgn} \left[ - \sum_{\alpha = 1}^m \frac{1}{M_\alpha} \left( \sum_{i = 1}^n \delta_{i,\alpha} X_i A_i \right)^2 \right] \neq +1~.
      \label{eq:3-12}
    \end{equation}

\section{Discussion}
\label{sec:5}
In Sec.~\ref{sec:3-3}, we proved that the FPSL cannot be created in any configuration of the single path. What prevents us from obtaining the FPSL? To understand the reason simply, let us consider the total RP fluctuations in a particular case.

\begin{figure}[htbp]
  \centering
    \includegraphics[clip,width=12.0cm]{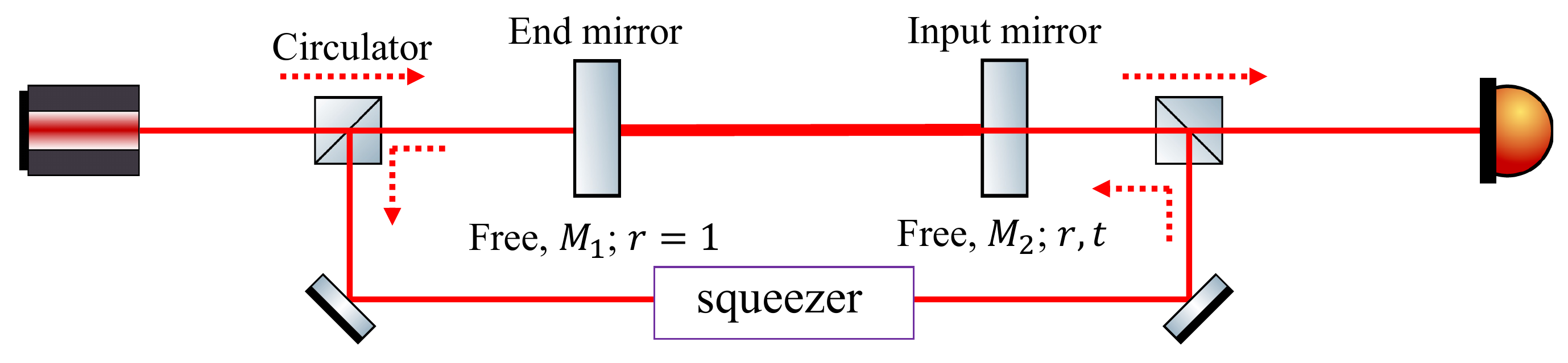}
  \caption{Sample geometry shared with the total-reflection end cavity mirror (left). This case is a single path case because the laser light is not divided in the whole path. The laser light is first reflected by the end mirror before being directed to the squeezer by a circulator. Using the circulator, the laser light is transmitted when incident from one direction and is reflected when incident from opposite direction. After passing through the squeezer, the laser light enters the Fabry-Perot cavity through the input mirror (right), and the reflected light is led to a photodetector by a circulator.}
  \label{fig:4-1}
\end{figure}

The geometry we consider is shown in Fig.~\ref{fig:4-1}. The laser light hits a free-mass mirror with $r = 1$ first. The laser beam is then squeezed by the crystal with the nonlinear polarization. Finally, the laser light is injected into a Fabry-Perot cavity. Its end mirror is the same as the mirror that the laser light hits at the first time; only the hitting direction of each laser differs. We assume that the laser light is resonant inside the cavity and that the cavity finesse $\mathcal{F}$ is high. This case is equivalent to the case with $m = 2,~n = 4,~l = 1$, and $n_1 = 2$ in Sec.~\ref{sec:3-3}. We will calculate $P_{\rm total}$ by denoting $\delta_{i,\alpha},~X_i,~A_i$. We consider each mirrors' mass as $M_\alpha$ in this geometry.

From Fig.~\ref{fig:3-2}, each squeezing effect with a nonlinear optical medium causes the $X_i$ to change. In this case, we use one squeezer; thus, the $X_i$ is described with Eq.~(\ref{eq:3-2}):
  \begin{align}
      X_i = \left\{
        \begin{array}{cl}
          X & (i = 1)~, \\
          \mathrm{e}^{r_\mathrm{sv}} X & (i = 2,3,4)~.
        \end{array} \right.
    \label{eq:4-1}
  \end{align}
The squeezer also changes the amplitude, the same as the amplitude fluctuations. On the process $i = 3$, the carrier amplitude is amplified inside the cavity. Using the initial carrier amplitude from a laser source $A$ and the amplitude after passing the squeezer $A'$, we can express the concrete values of $A_i$ from Eq.~(\ref{eq:2-19})
  \begin{align}
      A_i = \left\{
        \begin{array}{cl}
          A & (i = 1)~, \\
          A' & (i = 2)~, \\
          \left( \dfrac{1 + r}{t} \right)^2 A' & (i = 3)~, \\
          A' & (i = 4)~.
        \end{array} \right.
    \label{eq:4-2}
  \end{align}
Figure~\ref{fig:4-1} also shows whether or not and how the laser light hits the two mirrors. The $\delta_{i,\alpha}$ of each mirror, according to the definitions in Sec.~\ref{sec:3-3}, are given as follows:
  \begin{align}
      \delta_{i,1} = \left\{
        \begin{array}{cl}
          +1 & (i = 1) \\
          0 & (i = 2) \\
          -1 & (i = 3) \\
          0 & (i = 4)
        \end{array} \right.,~~
      \delta_{i,2} = \left\{
        \begin{array}{cl}
          0 & (i = 1,2) \\
          +1 & (i = 3) \\
          0 & (i = 4)
        \end{array} \right.~,
    \label{eq:4-3}
  \end{align}
where $\alpha = 1$ means the end mirror of the cavity and $\alpha = 2$ the input mirror. We can obtain the subtotal RP fluctuations by substituting Eqs.~(\ref{eq:4-1}) - (\ref{eq:4-3}) to Eq.~(\ref{eq:3-11}); that of the end mirror is written by
  \begin{equation}
      P_{{\rm total},\alpha = 1} = - \frac{1}{M_1} \cdot \frac{1}{\mathrm{e}^{r_\mathrm{sv}}} \left[ A - \left( \frac{1 + r}{t} \right)^2 \mathrm{e}^{r_\mathrm{sv}} A' \right]^2 X^2~,
    \label{eq:4-4}
  \end{equation}
and that of the input mirror is also
  \begin{equation}
      P_{{\rm total}, \alpha = 2} = - \frac{1}{M_2} \cdot \frac{1}{\mathrm{e}^{r_\mathrm{sv}}} \left[ \left( \frac{1 + r}{t} \right)^2 \mathrm{e}^{r_\mathrm{sv}} A' \right]^2 X^2~.
    \label{eq:4-5}
  \end{equation}
Summing up the two components, the total RP fluctuations are
  \begin{align}
      P_{\rm total} &= \sum_{\alpha = 1,2} P_{{\rm total}, \alpha} \notag \\
          &= - \frac{1}{\mathrm{e}^{r_\mathrm{sv}}} \left[ \frac{1}{M_1} \left\{ A - \left( \frac{1 + r}{t} \right)^2 \mathrm{e}^{r_\mathrm{sv}} A' \right\}^2 + \frac{1}{M_2} \left\{ \left( \frac{1 + r}{t} \right)^2 \mathrm{e}^{r_\mathrm{sv}} A' \right\}^2 \right] X^2~.
    \label{eq:4-6}
  \end{align}

The reason why no proper configuration to create the special squeezed light exists in the general single path can be explained in a similar way. We expand Eq.~(\ref{eq:4-6}) as an example:
  \begin{align}
      P_{\rm total} = - \frac{1}{\mathrm{e}^{r_\mathrm{sv}}} \left[ \frac{A^2}{M_1} + \left( \frac{1}{M_1} + \frac{1}{M_2} \right) \left( \frac{1 + r}{t} \right)^4 \mathrm{e}^{2r_\mathrm{sv}} {A'}^2 - \frac{2}{M_1} \left( \frac{1 + r}{t} \right)^2 \mathrm{e}^{r_\mathrm{sv}} A A' \right] X^2~.
    \label{eq:4-13}
  \end{align}
There is an opposite sign term (the third term). It indicates that the FPSL exists in the single path temporarily. From the calculation, the FPSL exists at the light path between before being injected into the cavity and after the first reflection in Fig.~\ref{fig:4-1}. However, it is compensated by the other conventional RP fluctuations. As a result, only the latter remains in the final beam. The same phenomenon occurs in other configurations; thus, the total RP fluctuations become Eq.~(\ref{eq:3-11}) and are not inverted.

\section{Conclusions}
We have suggested the possibility that the RP induced phase fluctuations in the light illuminating an interferometer with free-mass mirrors can be completely canceled by the FPSL. Then we searched for an optical configuration to create the FPSL in a single path. We considered the following optical processes: the light being injected into a cavity, hitting the free-mass mirrors normally or at an angle, and passing through a nonlinear optical device. We also considered combinations of these processes.

We have found that no suitable configuration to produce the FPSL existed no matter what combined the optical processes we used. This proof was discussed starting from the few special cases and finally concluding as a general case in Sec.~\ref{sec:3-3}. The reason why we cannot obtain the FPSL is that the FPSL, which can temporarily happen in the single path, is always compensated by the conventional phase fluctuations. We have shown this in Sec.~\ref{sec:5}.

In the next phase of this research, we will examine the configuration of dividing laser light by a BS in addition to the optical processes discussed in this paper. While the BS will enable us to expand the search space for the FPSL, we must consider vacuum fluctuations that are inevitably injected via the BS. We will investigate the possibility of a proper geometry to create the FPSL in the two (or more)-divided path cases.

\section*{Acknowledgments}
We would like to thank Stanley E. Whitcomb for English editing. This work was supported by Murata Science Foundation and JSPS KAKENHI, Grant Number JP19H01924.


\end{document}